\documentclass[aps,twocolumn,amsmath,amssymb,showpacs,floatfix,prb]{revtex4}
\usepackage{graphicx}

\begin{document}

\title{Linear Response Theory and the Universal Nature of the Magnetic
Excitation Spectrum of the Cuprates}
\author{M. R. Norman}
\affiliation{
Materials Science Division, Argonne National Laboratory, Argonne,
Illinois 60439}
\date{\today}
\begin{abstract}
Linear response theory, commonly known as the random phase approximation (RPA),
predicts a rich magnetic excitation spectrum for d-wave superconductors.  Many of
the features predicted by such calculations appear to be reflected in inelastic
neutron scattering data of the cuprates.  In this article, I will present results from RPA calculations
whose input is based on angle resolved photoemission data, and discuss possible relevance
to inelastic neutron scattering data of La$_{2-x}$Sr$_x$CuO$_4$ (LSCO), YBa$_2$Cu$_3$O$_{6+x}$
(YBCO), and Bi$_2$Sr$_2$CaCu$_2$O$_{8+x}$ (Bi2212)
in their superconducting and
non-superconducting phases.  In particular, the question of the universality of the magnetic
excitation spectrum will be addressed.
\end{abstract}

\maketitle

\section{Introduction}

The nature of the magnetic excitation spectrum of the cuprates is an important topic in its
own right.  But it has many implications beyond this as well.  It is of special relevance to those
theories that propose a `magnetic' origin for cuprate superconductivity \cite{AC}.  And it is of particular
interest to the current debate whether magnetic excitations or phonons are responsible
for certain strong coupling features observed in tunnelling, infrared  conductivity, and
angle resolved photoemission spectra \cite{RMP1,RMP2,davis}.

Although  there are a large number of  theories for explaining the observed magnetic
excitation spectrum,  to a first approximation, these theories can be collapsed into two
groups.  The first are linear response calculations based on a two dimensional Fermi surface
and the presence of a d-wave energy gap.  In this group are included calculations which
go beyond RPA (such as FLEX \cite{bickers}) and the generalization to the particle-particle
channel of the SO(5) approach \cite{SO5}.  The other group are those based on coupled
spin ladders \cite{uhrig,seibold}.  They assume phase segregation of the material into
undoped antiferromagnetic domains separated by one-dimensional `stripes' containing the doped
holes.  They have the advantage of treating the full quantum
mechanical nature of the spins, but have the disadvantage that fermionic degrees of freedom
are neglected.

It is not the intent here to discuss the relative merits of one approach versus the other.  In some
sense, both theories are different limits of a more complex theory which properly treats spin and charge
degrees of freedom.  Rather, I wish to discuss what success linear response calculations have
in regards to describing magnetic excitation properties of the cuprates,  but as well discuss their limitations.  In particular,
I wish to address the question of whether there is a `universal' behavior of the spin excitation
spectrum based on the RPA results.  Ultimately, I hope these results will have some relevance
to a final understanding of the magnetic behavior of these fascinating materials.

\section{Methodology}

The RPA expression for the interacting susceptibility is
\begin{equation}
\chi(q,\omega) = \frac{\chi_0(q,\omega)}{1 - U \chi_0(q,\omega)}
\end{equation}
In this equation,  $\chi_0$ is the polarization bubble constructed from
bare Greens functions, and $U$ is the effective Hubbard interaction which results from
projection onto the single band subspace (that is, the band that is the antibonding combination
of copper $d_{x^2-y^2}$ and oxygen $p_x$ and $p_y$ orbitals \cite{WP}).

It is often stated that RPA is a weak coupling approach, applicable only to
heavily overdoped materials which exhibit Fermi liquid like behavior \cite{JTrev}.  But it was shown
by Schrieffer, Wen, and Zhang \cite{SWZ} that RPA reproduces the superexchange
J and the resulting spin wave dispersion of the undoped material.  Obviously, there are important
quantum corrections to these results, but these do not impact the overall correctness of the approach.
One important point to remark is that the calculation of Ref.~\onlinecite{SWZ}
was self-consistent in the sense that the gapless
nature of the spin excitations at $q=(\pi,\pi)$ is a direct consequence of the mean-field equation for the Hubbard
gap.  Such self-consistency is difficult to implement in the doped case because of the added
complication of having to calculate the screening caused by the doped holes.

There are two ways to proceed for the doped case.  Continue to use bare Greens
functions and replace $U$ by an effective $U_{eff}$ which accounts for screening \cite{scal,vilk}.  Or
take the infinite $U$ limit, replace the Greens functions by those of the t-J model, and replace
$U$ by $J_{eff}$ where $J_{eff}$ takes into account the reduction of $J$ by the introduction of
doped holes (in the t-J approach, one
is forced to scale $J$ down to prevent a magnetic instability from occurring \cite{BLee}).  An important
additional remark is that the screened interaction is now dependent on transfered momentum, $q$.
This is particularly clear in the t-J model, where $J(q) = -J (\cos(q_xa)+\cos(q_ya))/2$ due to
exchange between near neighbor copper sites.

One limitation of the RPA approach is the use of bare Greens functions as opposed
to dressed ones.  But it is well known that using dressed Greens functions means that
vertex corrections must be included \cite{vilk}.  The neglect of such vertex corrections typically leads to
worse results than the RPA.

The approach taken in this paper will be somewhat more phenomenological.  The bare Greens 
functions will be based on dispersions derived from angle resolved photoemission.  The resulting
$U_{eff}$ will be taken as an adjustable constant, designed to reproduce the location of the S=1 resonance mode at $q=(\pi,\pi)$.  Calculations have been performed as well assuming $U_{eff}$
has the same q dependence as $J(q)$.  In a more complete theory, one might work instead 
within the context of dynamical mean field theory and its cluster generalization \cite{AMT}.

The heart of the RPA calculation turns out to be the structure of the bare polarization bubble,
$\chi_0$.  For a superconductor this is \cite{schrieff}
\begin{eqnarray}
\lefteqn{\chi_0(q,\omega) =} \nonumber \\
& & \sum_k\{\frac{1}{2}(1+\frac{\epsilon_k\epsilon_{k+q}+\Delta_k\Delta_{k+q}}
{E_kE_{k+q}})\frac{f(E_{k+q})-f(E_k)}{\omega-(E_{k+q}-E_k)+i\delta} \nonumber \\
& & +\frac{1}{4}(1-\frac{\epsilon_k\epsilon_{k+q}+\Delta_k\Delta_{k+q}}
{E_kE_{k+q}})\frac{1-f(E_{k+q})-f(E_k)}{\omega+(E_{k+q}+E_k)
+i\delta} \nonumber \\
& & +\frac{1}{4}(1-\frac{\epsilon_k\epsilon_{k+q}+\Delta_k\Delta_{k+q}}
{E_kE_{k+q}})\frac{f(E_{k+q})+f(E_k)-1}{\omega-(E_{k+q}+E_k)+i\delta}\}
\end{eqnarray}
where $E_k = \sqrt{\epsilon_k^2+\Delta_k^2}$.
The coherence factors play a critical role in the results \cite{pwa}.  For an s-wave
superconductor, the coherence factor (in term 3 of Eq.~2) vanishes on the Fermi surface.  The net result
is that there are no spin collective modes in this case.  On the other hand, for
an order parameter satisfying the condition $\Delta(k+q)=-\Delta(k)$, the coherence
factor becomes maximal on the Fermi surface (equal to 2).  As a consequence, the imaginary
part of the bubble has a step discontinuity at threshold, and thus the real part has
a logarithmic divergence.  This divergence guarantees the appearance of a collective mode
below threshold.  This collective mode energy at $q=(\pi,\pi)$ will be denoted as
the resonance energy.

There is a common misconception that the appearance of the collective mode
requires the presence of a d-wave superconducting gap.  In fact, the only
requirement is that the coherence factors do not vanish on the Fermi surface.
As an example, imagine that the pseuodgap phase is a phase disordered version
of a superconductor.  To a first approximation, we can then set 
$<\Delta_k\Delta_{k+q}>$=0 \cite{millis}.  In this case, the coherence factor
is now unity on the Fermi surface independent of the value of $k$, and one finds
a collective mode below threshold as in the case of a d-wave superconductor.  As
the coherence factor is now half that of the superconductor, then
larger values of $U_{eff}$ are needed to obtain similar collective mode
energies.  In this paper,  a d-wave order parameter of the form 
$\Delta(k) = \Delta_0 (\cos(k_xa)-\cos(k_ya))/2$ is assumed, although for some calculations
the phase disordered approximation was invoked as well.

Finally, some technical remarks.  Eq.~2 is solved by replacing $\delta$ by some finite
$\Gamma$ and doing a simple sum over the Brillouin zone.  The smaller $\Gamma$ is,
the more k points are needed in the sum.  Results have been generated for $\Gamma$ ranging
between 0.1 and 2 meV, with zone meshes ranging from 400 by 400 to 4000 by 4000, but
in this paper, results are shown for $\Gamma$=2 meV (400 by 400 mesh).
All calculations were performed for a temperature of 1 meV.  All susceptibilities
quoted here are states per eV per CuO$_2$ formula unit and should be multiplied by the matrix element
$\sum_{\sigma}g^2\mu_B^2\langle\sigma|S_z|\sigma\rangle^2=2\mu_B^2$ to compare
to experiment.

\section{Results - Normal State}

The most important input into the RPA calculations is the assumed form of the fermionic
dispersion.  For instance, a well studied case is a tight binding model with just near
neighbor hopping \cite{SO5}.  When considering the response at $q=(\pi,\pi)$, the
threshold for the polarization bubble corresponds to exciting from the node of the d-wave
order parameter, $k_N$, to an unoccupied state, $k_N+q$, along the zone diagonal.
As such, the threshold is independent of the d-wave energy gap.
The Fermi surfaces observed by angle resolved photoemission, though, differ considerably
from those predicted by a near neighbor tight binding model.  In particular, the observed
Fermi surfaces are characterized by `hot spots', that is, points on the Fermi surface which
satisfy the condition $\epsilon_k = \epsilon_{k+q} = 0$.  In this case, the threshold of
the bubble at $q=(\pi,\pi)$ is determined by twice the energy gap at the hot spots.

\begin{figure}
\centerline{\includegraphics[width=3.4in]{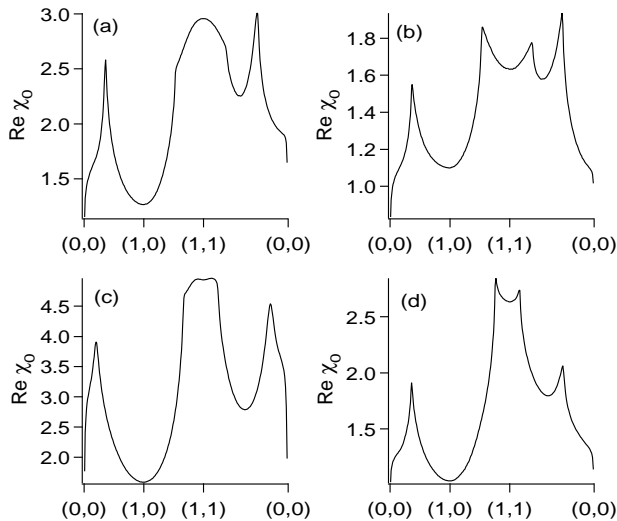}}
\caption{Re $\chi_0$ in the normal state at zero energy along the symmetry axes of the zone
for (a) tb1 (b) tb2 (c) tb3 and (d) tb4.  The zone notation is in $\pi$ units.}
\label{fig1}
\end{figure}

\begin{figure}
\centerline{\includegraphics[width=3.4in]{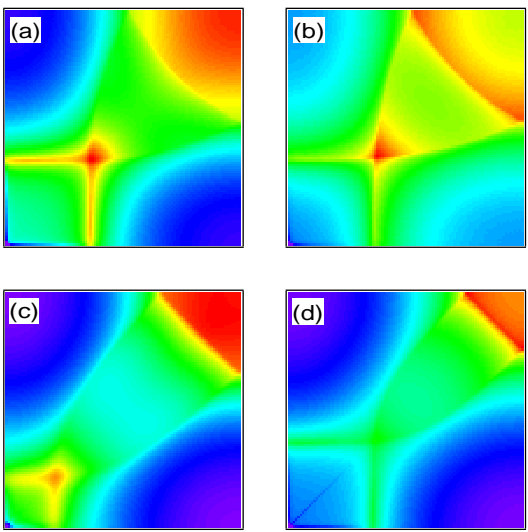}}
\caption{2D zone plots of Fig.~1.  The lower left corner is at (0,0) and the upper right
corner is at $(\pi,\pi)$.  The dynamic range is the same as in Fig.~1.}
\label{fig2}
\end{figure}

When considering the response for a general $q$, there is even more sensitivity
to the assumed  fermionic dispersion.  To illustrate this point, in Fig.~1,
I show the real part of the normal state ($\Delta=0$) bubble at $\omega=0$ for four different
dispersions along the three symmetry axes of the 2D Brillouin zone.
The first dispersion (tb1) is that based on an early tight binding fit to ARPES spectra
on Bi2212 \cite{mike95}.  Two prominent features are observed (Fig.~1a).  First, there is a broad
maximum around $q=(\pi,\pi)$.  Second, there are incommensurate peaks along the bond
and diagonal directions which correspond to scattering between the antinodal sections of
the Fermi surface near the $(\pi,0)$ points.  As can be seen from the 2D plot in Fig.~2a,
this scattering forms a square box around the zone center, with maximum intensity at the
box corners.  These peaks in the zero $\Gamma$ limit are
logarithmically divergent \cite{wang}. Although potentially important for the charge response,
they are probably not so for the spin response (since $J(q)$ has opposite sign in this
region of the zone).

The other noticeable feature of Fig.~1a is the broad maximum around $q=(\pi,\pi)$.  As
a consequence, in the gapped case, the dominant behavior below the resonance energy is 
commensurate \cite{mike1,mike2}.  But this behavior is actually the exception, rather than the
rule.  At this point, a discussion on how the fermionic dispersion is chosen is in order \cite{mike95}.
A fit is performed using a six parameter tight binding model which includes real space lattice
vectors of type (0,0), (1,0), (1,1), (2,0), (2,1), and (2,2).
In the original analysis, the fitting variables were
(1) the position of the Fermi surface along $(0,0)-(\pi,\pi)$ (the node), (2) the Fermi velocity at
the node, (3) the Fermi surface along  $(\pi,0)-(\pi,\pi)$ (the antinode), (4) the energy at $(\pi,0)$,
(5) the curvature of the dispersion at $(\pi,0)$ (along the $k_x$ direction), and (6) the energy
at $(\pi,\pi)$ (this last condition is simply for stability of the fit).  The behavior around $q=(\pi,\pi)$
in Fig.~1a is largely a consequence of the fact that the saddle point energy at $(\pi,0)$, -34 meV,
is close
to the Fermi energy.  For dispersions where this is displaced deeper in energy, a minimum is found
in Re $\chi_0$ at $q=(\pi,\pi)$ rather than a maximum.

To illustrate this, a tight binding analysis has been performed on the latest high resolution ARPES
measurements on optimal doped Bi2212 \cite{adam05}.  These data reveal an isotropic Fermi velocity with a value of
1.5 eV$\AA$.  Fixing this condition forces the energy at $(\pi,0)$ to lie at -119 meV, as
compared to -34 meV in the earlier fit \cite{bifoot}.  This difference (dispersion tb2) has a profound effect on the real 
part of
the bubble, as can be seen in Fig.~1b.  Now,  incommensurate peaks are observed around
$q=(\pi,\pi)$ as well, and as can be seen in Fig.~2b, they form a diamond shaped pattern.  This
structure is actually discernible in Fig.~1a as kinks, but it is obscured by the dominant 
maximum around
$q=(\pi,\pi)$.  The behavior of Fig.~1b is also found for the t,t$^{\prime}$ dispersion used in 
Ref.~\onlinecite{gang5}, as the $(\pi,0)$ point in that case is at -129 meV.

\begin{figure}
\centerline{\includegraphics[width=3.4in]{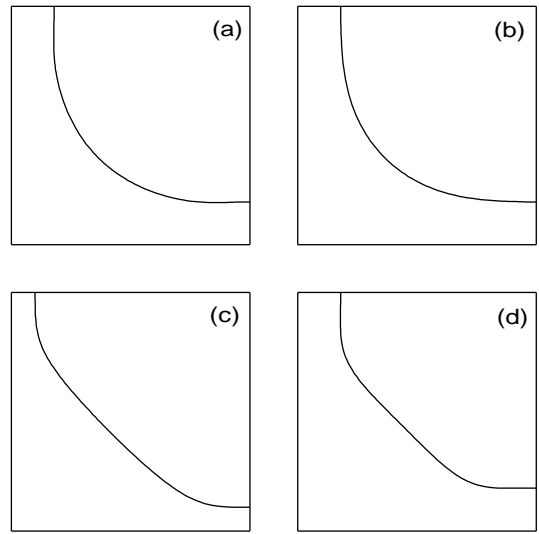}}
\caption{Fermi surfaces corresponding to Fig.~1.
The lower left corner is at (0,0) and the upper right
corner is at $(\pi,\pi)$.}
\label{fig3}
\end{figure}

The incommensurate effects can also be strengthened by flattening the Fermi surface around
the node \cite{BLee,mike1}.  A dispersion which incorporates this effect, as well as the presence
of a flat quasiparticle band near $(\pi,0)$, has the Fermi surface illustrated in Fig.~3c
(this is dispersion tb3 of Ref.~\onlinecite{mike2}, and is denoted as tb3 here as well).
In this case, the effect of the enhanced
incommensurability and the tendency to have a maximum at $q=(\pi,\pi)$ due to the small energy of the $(\pi,0)$ point
(-10 meV) approximately balance, leading to a net flat behavior around $q=(\pi,\pi)$, as can be
seen in Fig.~1c.

One issue is that there is not much inelastic neutron scattering data on Bi2212 because of the
small sample sizes, though this is starting to change.\cite{fauque}
A large amount of data is present for YBCO, but the electronic structure
of this material is complicated because of its strong orthorhombicity and the presence of metallic
chain layers.  Instead, I consider LSCO.  This material has the advantage of having extensive
inelastic neutron scattering data and a simpler electronic structure (no bilayer splitting).
Recently, Tranquada and co-workers \cite{tranq04} have pointed out that the inelastic magnetic 
response of non-superconducting La$_{1.875}$Ba$_{0.125}$CuO$_4$ (LBCO) is very similar to that of
superconducting YBCO \cite{hayd04}, which makes this material of particular interest.  Extensive ARPES data
exist on LSCO.  Recently, an in depth study was done on underdoped LSCO \cite{zhou} and
revealed a quite striking Fermi surface (Fig.~3d).  The Fermi surface is characterized by
two straight sections, one centered about the node, the other about the antinode.  This Fermi
surface, and the observed dispersion (denoted as tb4), 
is fit using the condition (1) node at (0.44,0.44)$\pi$, (2) velocity at the
node (1.8 eV$\AA$), (3) curvature of the Fermi surface at the node (zero), 
(4) antinode at (1,0.18)$\pi$, (5) $(\pi,0)$
energy at -70 meV, and (6) $(\pi,\pi)$ energy at 1 eV (for fit stability).  The result for the zero
frequency real response is shown in Figs.~1d and 2d.  Again, note the strong ``box-like"
structure around the zone center, and the pronounced incommensurate behavior around
$(\pi,\pi)$ due to the flat Fermi surface near the node.  In this context, it should be noted
that the box-like structure near $q=(0,0)$ is due to nesting of the antinodal regions,
but the diamond-like structure near $q=(\pi,\pi)$ is due to nesting of the nodal regions.
This is in contrast with the alternate stripes model, where the incommensurability around
these two $q$ vectors have the same origin.

\begin{table}
\caption{Tight binding dispersions based on angle resolved photoemission data.
The first four columns list the coefficient, $c_i$, of each term (eV), that is
$\epsilon(\vec k) = \sum c_i \eta_i(\vec k)$.
The last column lists the basis
functions (the lattice constant $a$ is set to unity).  Below this are several properties
of these dispersions - the node, $(k_N,k_N)$, the antinode, $(1,k_{AN})$ (in $\pi/a$ units),
the nodal velocity, $v_N$, the antinodal velocity, $v_{AN}$ (in eV$\AA$),
the energy at $(\pi,0)$, $\epsilon_M$ (eV), the bandwidth, W (eV), and the value
of $U$ needed for resonance at 40 meV in the superconducting state.
}
\begin{ruledtabular}
\begin{tabular}{rrrrc}
tb1 & tb2 & tb3 & tb4 & $2\eta_i(\vec k)$ \\
\colrule
 0.1305 & 0.1960 & 0.1197 & 0.0801 & $2$ \\
-0.5951 &-0.6798 &-0.5881 &-0.7823 & $\cos k_x + \cos k_y$ \\
 0.1636 & 0.2368 & 0.1461 & 0.0740 & $2\cos k_x \cos k_y $ \\
-0.0519 &-0.0794 & 0.0095 &-0.0587 & $\cos 2 k_x + \cos 2 k_y$ \\
-0.1117 & 0.0343 &-0.1298 &-0.1398 &  $\cos 2k_x\cos k_y+\cos k_x\cos 2k_y$ \\
 0.0510 & 0.0011 & 0.0069 &-0.0174 & $2\cos 2k_x \cos 2k_y $ \\
 \colrule
 0.365 & 0.365 & 0.414 & 0.440 & $k_N$ \\
 0.180 & 0.180 & 0.100 & 0.180 & $k_{AN}$ \\
 1.63 & 1.50 & 1.14 & 1.77 & $v_N$ \\
 0.48 & 1.50 & 0.24 & 0.91 & $v_{AN}$ \\
 -0.034 & -0.119 & --0.010 & -0.070 & $\epsilon_M$ \\
 1.414 & 1.291 & 1.436 & 1.844 & W \\
 0.274 & 0.563 & 0.214 & 0.332 & U \\
\end{tabular}
\end{ruledtabular}
\end{table}

A summary of the properties of these four tight binding dispersions are listed in
Table 1.  A number of other tight binding dispersions have been analyzed as well,
including some that account for bilayer splitting \cite{bilayer}.  The results fall within the
range of behavior discussed for the four dispersions here.  In addition, bilayer splitting
leads to some extra details in the even (optic) spin response due to differing momentum
locations of features
from bonding-bonding and antibonding-antibonding responses \cite{eremin2}.  For simplicity, these
details are not discussed here as they do not occur in the dominant odd (acoustic)
spin channel which involves only the bonding-antibonding response.

\section{Results - Superconducting State}

The presence of a d-wave energy gap leads to a polarization bubble which now becomes gapped
for all $q$ vectors in the first zone but two (these vectors being the ones that connect the nodes
of the d-wave order parameter).   Because of the strong anisotropy of the d-wave gap, the lower
edge of the particle-hole continuum also has a strong $q$ dependence.  This continuum edge
is plotted along the zone symmetry axes in Fig.~4 for the four dispersions studied here.  The
threshold at $(\pi,\pi)$ corresponds to twice the energy gap at the hot spots.  As one moves away
from this wavevector, the threshold splits \cite{schnyder,gang5} into several ones:  two for $\vec{q}$
along $(\pi,x)$, and three for $\vec{q}$ along $(x,x)$.  Along $(x,x)$, the zero threshold at 
$\vec{q}_N=2\vec{k}_N$ obviously corresponds to two nodal points, the analogous minimum
along $(\pi,x)$ involving only one nodal point (the other being at $\vec{k}_N+(\pi,x)$).

\begin{figure}
\centerline{\includegraphics[width=3.4in]{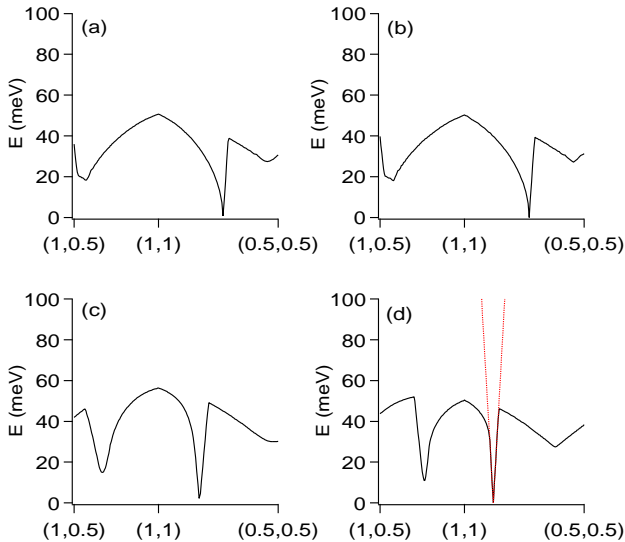}}
\caption{Particle-hole continuum edge along the symmetry axes of the zone.  The
maximum d-wave gap is 30 meV.  The dashed line in (d) corresponds to the process
involving $k_N$  and $k_N+q$.  The zone notation is in $\pi$ units.}
\label{fig4}
\end{figure}

As briefly discussed above, the effect of an energy gap has a profound impact on the
polarization bubble.  The presence of an energy gap, along with the fact that the coherence
factors at threshold do not generally vanish in the d-wave case, leads to a step jump in the imaginary
part of the bubble at threshold.  By Kramers-Kronig, such a step jump causes a logarithmic
divergence in the real part of the bubble at threshold.  As a consequence, any finite value
of $U$ will lead to the presence of a collective mode below threshold.  The exception is
in those cases where $\Delta_k$ and $\Delta_{k+q}$ have the same sign, which in the d-wave
case can occur
for $q$ vectors significantly displaced from $(\pi,\pi)$.  In addition, for a pole to
occur, then $U$ must have a positive sign (this is relevant to near-neighbor exchange
models, where the positive sign only occurs for $q_x+q_y > \pi$).

The value of $U$ needed to have a pole at $q=(\pi,\pi)$ at a particular energy
varies significantly among the dispersions
looked at here (the values for 40 meV are also listed in Table 1).  This is a combination of the different Fermi
surfaces shown in Fig.~3, the different energies of the van Hove singularity at $(\pi,0)$ for these 
dispersions, and the different overall bandwidths.  This variation of $U$ involves a rather subtle issue concerning
whether one uses a bare dispersion based on band theory, or a renormalized (quasiparticle)
dispersion, for $\epsilon_k$ in Eq.~2 (and also the issue of bilayer splitting).  Again, the purpose 
here is to simply look at a range of dispersions and contrast their behaviors.

Figs.~5-8 show the variation of the imaginary part of the RPA susceptibility (Eq.~1) with $U$
tuned to yield a pole at $q=(\pi,\pi)$ at 40 meV (50 meV for dispersions tb2 and tb4 - the
reason for this difference will be discussed below).  $U$ was treated as a constant, though
calculations were also performed for $U=J(q)$.   A maximum energy gap of 30 meV was
assumed.

\begin{figure}
\centerline{\includegraphics[width=3.4in]{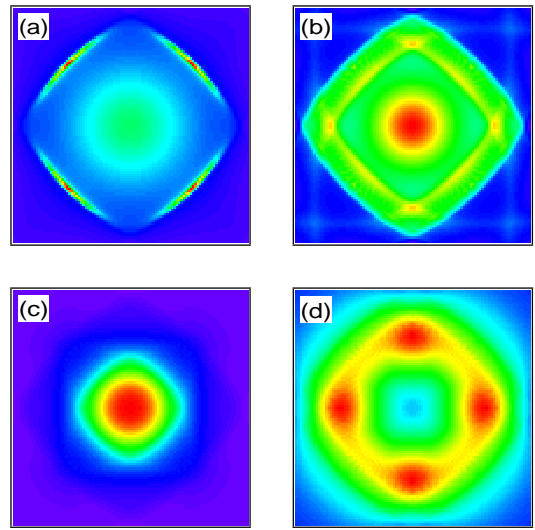}}
\caption{2D zone plots of Im$\chi$ for tb1 at (a) 10 meV (b) 30 meV (c) 40 meV
and (d) 50 meV.  The maximum d-wave gap is 30 meV, and $U$ (0.274 eV) is chosen to
give a resonance at 40 meV.  In Figs.~5-8, the lower left corner is at $(\pi/2,\pi/2)$, the
center at $(\pi,\pi)$.  Scale is such that the lowest is 0 and the highest is (a) 0.7 (b) 4.4
(c) 112 and (d) 21.3}
\label{fig5}
\end{figure}

Fig.~5 shows results for tb1.  At low energies, one sees weak intensity near $\vec{q}_N$ corresponding to the nodes.
As the energy is raised to exceed the minimum threshold along $(\pi,x)$, this incommensurate pattern 
rotates 45 degrees.  The preference for this rotated pattern was discussed by Schulz \cite{shulz,BLee}
and is a consequence of the fact that translation by $(x,x)$ only brings the Fermi surface into
coincidence in one quadrant of the zone,
but translation by $(\pi,x)$ brings it into coincidence in two quadrants (Fig.~9).
In addition, though, one sees a global maximum at $q=(\pi,\pi)$ which is  a consequence of the
global maximum seen in the normal state response at zero energy discussed earlier.  Above
the resonance energy (40 meV), the response becomes incommensurate, with the displacement of
the maxima away from $(\pi,\pi)$ increasing in magnitude (but the response decreasing in strength) with increasing energy.  For a range of energies above resonance ($\sim$ 12 meV), the incommensurate response corresponds to damped poles, and mostly has maxima along along $(\pi,x)$,
but for some energies the maxima are along $(x,x)$.  This is due to the anisotropy
of the splitting of the particle-hole threshold as one moves away from $(\pi,\pi)$.  The overall
response is best appreciated by plotting the intensity as a function of energy along the (x,x)
direction in momentum  space as shown in Fig.~10a.  In the pole region (40 - 52 meV), one sees a magnon-like (quadratic) dispersion.  Above this region, the response rapidly loses strength and the dispersion becomes more steep.

\begin{figure}
\centerline{\includegraphics[width=3.4in]{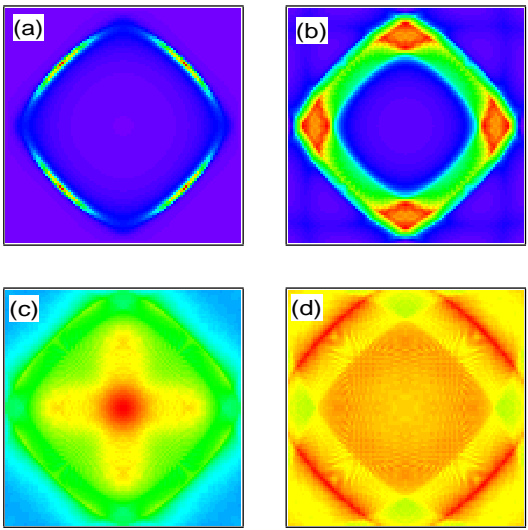}}
\caption{2D zone plots of Im$\chi$ for tb2 at (a) 10 meV (b) 30 meV (c) 50 meV
and (d) 100 meV.  The maximum d-wave gap is 30 meV, and $U$ (0.508 eV) is chosen to
give a resonance at 50 meV.  Scale is such that the lowest is 0 and the highest is (a) 6.3 (b) 20.6
(c) 22.6 and (d) 7.0}
\label{fig6}
\end{figure}

I now contrast this behavior with that from the next dispersion, tb2.  This dispersion, based on
ARPES fits to the bonding Fermi surface in Bi2212, has (1) an isotropic velocity around the Fermi
surface and (2) an energy for the $(\pi,0)$ point which is much deeper than for tb1 (-119 meV
as compared to -34 meV).  As a consequence, this dispersion yields a bubble whose real part
has a relatively weak momentum dependence, and is very typical of many of the dispersions used in
the theoretical literature.  In fact, the momentum dependence is so weak,
that with a constant $U$, the condition to yield a resonance at 40 meV at $(\pi,\pi)$ implies the
presence of long range order (that is, there is a $q$ value at zero energy where the real part
of the bubble exceeds 1/$U$).  This forces us to move the resonance condition close to the edge
of the continuum (where the real part of the bubble has a peak) in order to avoid this problem.
And obviously, because of the weak momentum dependence, the value of $U$ needed to obtain
a resonance condition is significantly larger than for tb1.  The results are plotted in Figs.~6 and 10b.
Below resonance, despite the lack of a commensurate response, the incommensurate response
is similar to tb1, being dominated by node-node processes at low energies, and then rotating
to a bond centered response once the minimum of the continuum edge along $(\pi,x)$ is exceeded.
The response becomes commensurate at resonance (50 meV), then one has a weak incommensurate
response above resonance whose maxima generally sit along the diagonals.  Although this
45 degree rotation of the incommensurate maxima is indeed what is seen in YBCO \cite{hayd04}
and LBCO \cite{tranq04}, the incommensurate response above resonance still has an overall
diamond shape to it unlike experiment.  This shape is a consequence of 2$k_F$ scattering which
can be seen as well in the normal state (Fig.~2b).  The  overall behavior is best appreciated in
Fig.~10b, where in contrast to dispersion tb1, a very clear `reverse magnon' dispersion is
evident for the resonance mode.  Above resonance, the incommensurate response is almost
dispersionless, and seems to follow the steep particle-hole response involving scattering along
the zone diagonal (dashed line in Fig.~4d).  This steepness is due to the high Fermi velocity along the nodal direction and
unlike the rest of the continuum response, is independent of the energy gap.

\begin{figure}
\centerline{\includegraphics[width=3.4in]{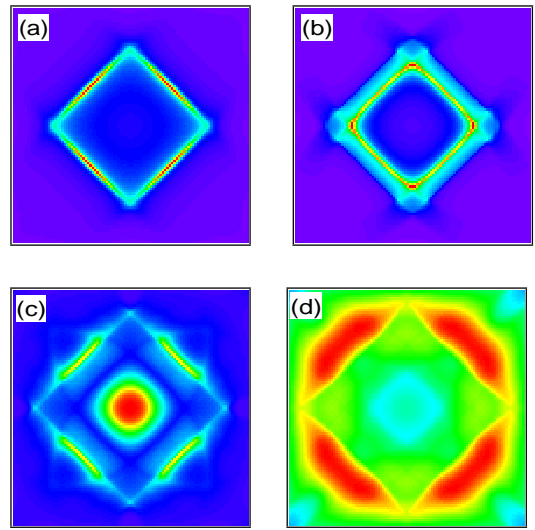}}
\caption{2D zone plots of Im$\chi$ for tb3 at (a) 10 meV (b) 30 meV (c) 40 meV
and (d) 50 meV.  The maximum d-wave gap is 30 meV, and $U$ (0.214 eV) is chosen to
give a resonance at 40 meV.  Scale is such that the lowest is 0 and the highest is (a) 4.0 (b) 101
(c) 112 and (d) 10.6}
\label{fig7}
\end{figure}

The next dispersion, tb3, had been designed to enhance the incommensurability by flattening
the Fermi surface around the node.  In Fig.~7, one can see the strong diamond-shaped incommesurate
pattern below resonance that is a result of the enhanced Fermi surface nesting in the nodal region.
The  evolution of the pattern is again a diagonal response at very low energies due to node-node
processes, a rotation by 45 degrees to a bond centered pattern once the threshold along $(\pi,x)$
is exceeded, commensurability at resonance (40 meV), and then again a complicated incommensurate
response above resonance due to the anisotropic splitting of the continuum edge as one
moves away from $(\pi,\pi)$.   The overall dispersion of the magnetic response is most visible
in Fig.~10c.  Note the pronounced downward dispersion of the resonance which represents a true
pole in the RPA susceptibility in this case (this response is below the continuum edge). 
One then sees a weak intensity gap near the node-node vector due to crossing into the continuum
(the so-called silent band \cite{pailhes}).  Beyond this, a new pole appears \cite{gang5} on the other side of the continuum (the edge of which is again controlled by the high Fermi velocity along the nodal
direction).  One then reenters the continuum, and the response rapidly loses intensity and the
damped pole-like response is lost above 55 meV.

I now turn to the last dispersion, tb4.  This is based on a fit to underdoped LSCO ARPES
data \cite{zhou}.  The Fermi surface in this case is characterized by strong nesting, both in the nodal
region, and also the antinodal one.  Putting the resonance condition at 40 meV yields the same
reversed magnon dispersion, silent band effect around $q_N$ (node-node vector) , and second mode behavior for $q < q_N$ that was so prominent for tb3 (Fig.~11a).  The one contrast is at higher energies, the incommensurability is much better defined than for
the other dispersions due to the strong nesting (this high energy dispersion follows the steep
dispersion along the nodal direction shown in Fig.~4d).

But for most of the results presented here, I choose to show instead the case where the resonance
condition is at 50 meV.  This is interesting for two reasons: (1) this energy is where the maximum
response was seen in LBCO \cite{tranq04}, and (2) this energy corresponds to the continuum edge
at $(\pi,\pi)$ (and thus where there is a maximum in the real part of the bubble).  $U$ in
this case is adjusted so that it is equal to the inverse of this maximum value.

\begin{figure}
\centerline{\includegraphics[width=3.4in]{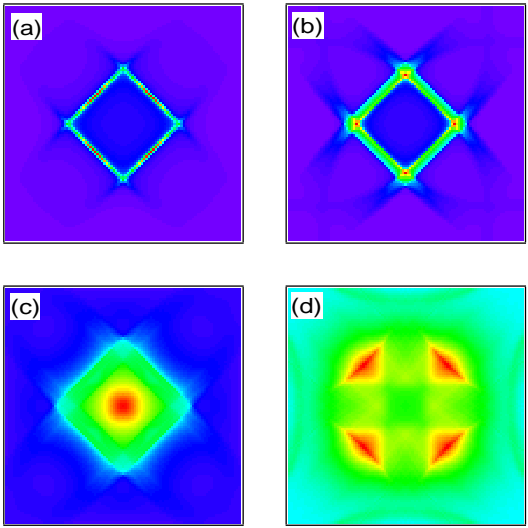}}
\caption{2D zone plots of Im$\chi$ for tb4 at (a) 10 meV (b) 30 meV (c) 50 meV
and (d) 100 meV.  The maximum d-wave gap is 30 meV, and $U$ (0.276 eV) is chosen to
give a resonance at 50 meV.  Scale is such that the lowest is 0 and the highest is (a) 1.4 (b) 10.6
(c) 27.1 and (d) 5.5}
\label{fig8}
\end{figure}

\begin{figure}
\centerline{\includegraphics[width=3.4in]{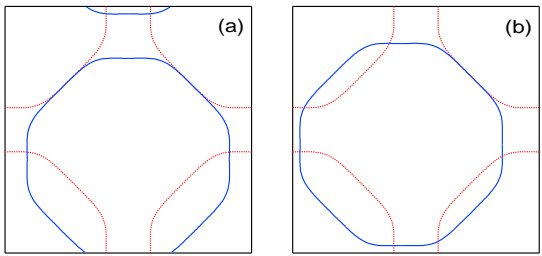}}
\caption{Fermi surface for tb4 (dashed line) and its $q$ translated image (solid line)
 for (a) $q=(1,.76)\pi$
and (b) $q=(.88,.88)\pi$.  Lower left corner is at $(-\pi,-\pi)$ and upper right at $(\pi,\pi)$.}
\label{fig9}
\end{figure}

\begin{figure}
\centerline{\includegraphics[width=3.4in]{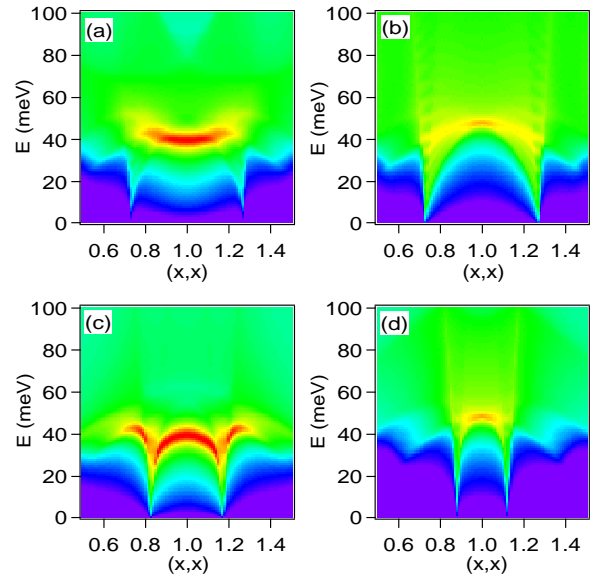}}
\caption{Plots of Im$\chi$ versus energy and $(x,x)\pi$ for (a) tb1 (b) tb2 (c) tb3
and (d) tb4.  Same conditions as Figs.~5-8.
The intensity is on a logarithmic grid ranging from 0.1 to 100.}
\label{fig10}
\end{figure}

\begin{figure}
\centerline{\includegraphics[width=3.4in]{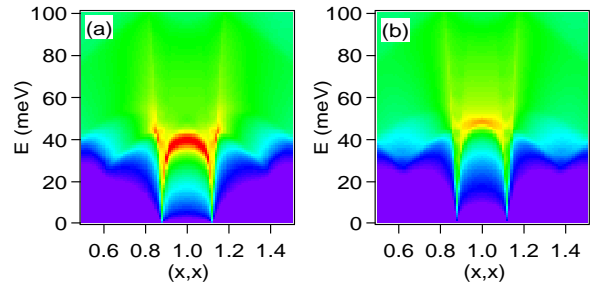}}
\caption{Plots of Im$\chi$ versus energy and $(x,x)\pi$ for tb4 for (a) superconducting
state with resonance at 40 meV (U=0.332 eV) and (b) pseudogap state with resonance at 
50 meV (U=0.343 eV).
The intensity is on a logarithmic grid ranging from 0.1 to 100.}
\label{fig11}
\end{figure}

The results are shown in Fig.~8 and Fig.~10d.  In this case (as well as when the resonance
is at 40 meV), the incommensurate response below resonance
is much cleaner than for the other dispersions, with very clear maxima along the $(\pi,x)$ direction once the continuum edge energy along this direction is exceeded, with
$x$ at the lowest energies being given by the condition $\vec{k}_N=(\pi/2-x/4,\pi/2-x/4)$. 
One again sees a clear rotation of the pattern by 45 degrees as
one crosses through the commensurate resonance energy, but as with tb2, the response above
resonance has an overall diamond shape.  Unlike the other dispersions, a well defined incommensurate
pattern with maxima along the diagonal direction persists to much higher energies, remaining clearly
visible up to 140 meV.  Above this energy, the response is less well defined, but still incommensurate.
Note the very well defined incommensurability at 100 meV, which is similar to that of the real part of the bubble (indicating pole-like behavior).
The interesting point about this dispersion, as commented above, is that the high energy response 
typically follows the outer branch of the dispersive response corresponding to scattering processes
along the nodal direction (dashed line in Fig.~4d).  This nodal velocity is set by the renormalized
Fermi velocity along the node, which in RVB type models is proportional to the superexchange, $J$.
This velocity is also essentially constant with doping \cite{zhounat}, suggesting a connection between
the universal behavior observed in both ARPES and INS data.

For this dispersion, calculations have also been performed in the pseudogap approximation.  This is
shown in Fig.~11b, and as can be seen, are virtually identical to the superconducting results
shown in Fig.~10d.  This indicates that the d-wave phasing relation is not necessary to obtain
the results shown in this paper (just an energy gap - though we again note that for the s-wave
superconducting case, there is no resonance effect).  In this connection, most of the dramatic
findings in INS data are in the underdoped regime where a pseudogap persists to very high
temperatures.

\begin{figure}
\centerline{\includegraphics[width=3.4in]{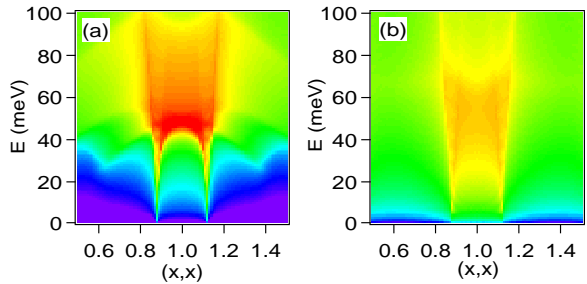}}
\caption{Plots of Im$\chi$ versus energy and $(x,x)\pi$ for tb4 for (a) superconducting state
and (b) normal state.  U is 0.276 eV.
The intensity is on a logarithmic grid ranging from 0.02 to 20.}
\label{fig12}
\end{figure}

\begin{figure}
\centerline{\includegraphics[width=3.4in]{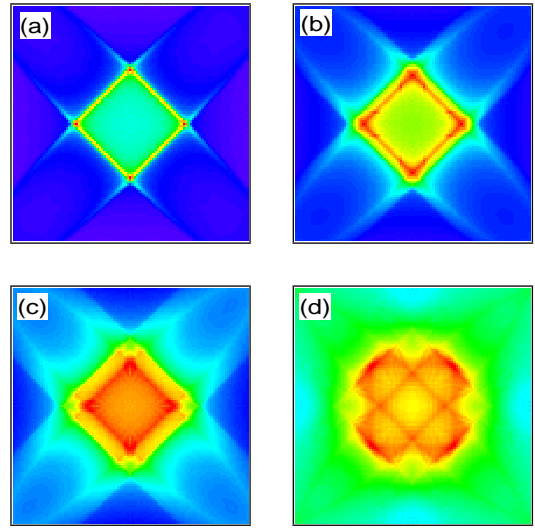}}
\caption{2D zone plots of Im$\chi$ for tb4 in the normal state at (a) 10 meV (b) 30 meV (c) 50 meV
and (d) 100 meV.  $U$ is 0.276 eV.  Scale is such that the lowest is 0 and the highest is (a) 4.8 (b) 6.8 (c) 7.2 and (d) 3.8}
\label{fig13}
\end{figure}

It is also amusing to present results for this dispersion when the energy gap is set to zero.  This is
shown in Fig.~12b for the same $U$ as used in the superconducting case in Fig.~10d.  The
response is always incommensurate, and is bond oriented for $\omega < \omega_c$ and
diagonal oriented for $\omega > \omega_c$ (Fig.~13).  $\omega_c$ is the energy where the
constant energy contour for the dispersion switches from being hole-like to electron-like.  Besides
the lack of a commensurate resonance response, and a spin gap due to the energy gap,
there are interesting similarities with the superconducting results.

\section{Conclusions}

In conclusion, RPA calculations show a remarkably rich behavior, in particular in the evolution
of the momentum response as a function of energy.  This behavior becomes even richer in
the superconducting and pseudogap phases due to the strong momentum anisotropy of
the particle-hole continuum edge due to the anisotropy of the energy gap.  There is a striking
similarity of many of these findings to experimental INS data in LSCO, YBCO, and Bi2212 -
in particular the reversed magnon behavior of the resonance mode, and the finding for
several dispersions of a rotation by 45 degrees of the incommensurate response when passing
through the resonance energy, which argue for universality in the magnetic response
as has been commented on in regards to experimental INS data.

On the other hand, there are some differences that need to be kept in mind.
For instance, the incommensurate response above resonance is generally more variable in the
calculations than indicated by experiment.  The fact that this is apparent as well in calculations
where the bare Greens functions are replaced by experimental Greens functions \cite{utpal}
indicates that this is a general issue.  The author speculates that the RPA calculations and
alternate ones based on coupled spin ladders are different limits of a more complete theory that
properly includes the full quantum mechanical nature of both the spin and charge degrees
of freedom.  The development of such a theory should help shed more light on the 
relation of the magnetic fluctuation spectrum and the existence of d-wave
superconductivity.

\acknowledgments
I would like to thank Andrey Chubukov, Steve Hayden, and John Tranquada for discussions, and
a special thanks to Herb Mook for his encouragement to initiate this project back in 1999.
The work was supported by the U. S. Dept. of Energy, Office of Science,
under Contract No. DE-AC02-06CH11357.


\begin{thebibliography}{99}

\bibitem{AC}
A. V. Chubukov, D. Pines, and J. Schmalian, {\it The Physics of 
Superconductors}, Vol. 1, ed. K. H. Bennemann and J. B. Ketterson (Springer,  Berlin, 2003), p. 495.

\bibitem{RMP1} A. Damascelli, Z. Hussain, and Z.-X. Shen,
Rev. Mod. Phys. {\bf 75}, 473 (2003).

\bibitem{RMP2}
D. N. Basov and T. Timusk,
Rev. Mod. Phys. {\bf 77}, 721 (2005).

\bibitem{davis}
J. Lee, K. Fujita, K. McElroy, J. A. Slezak, M. Wang, Y. Aiura, H. Bando, M. Ishikado, T. Matsui,
J.-X. Zhu, A. V. Balatsky, H. Eisaki, S. Uchida, J. C. Davis, Nature (London) {\bf 442}, 546 (2006).

\bibitem{bickers}
N. E. Bickers, D. J. Scalapino, S. R. White, Phys. Rev. Lett. {\bf 62}, 961 (1989).

\bibitem{SO5}
E. Demler, W. Hanke and S. C. Zhang, 
Rev. Mod. Phys. {\bf 76}, 909 (2004).

\bibitem{uhrig}
G. S. Uhrig, K. P. Schmidt and M.
Gruninger, Phys. Rev. Lett. {\bf 93}, 267003 (2004).

\bibitem{seibold}
G. Seibold and J. Lorenzana,
Phys. Rev. Lett. {\bf 94}, 107006 (2005).

\bibitem{WP}
W. E. Pickett, Rev. Mod. Phys. {\bf 61}, 433 (1989).

\bibitem{JTrev}
J. T. Tranquada, cond-mat/0512115 (2005).

\bibitem{SWZ}
J. R. Schrieffer, X. G. Wen and S. C. Zhang, Phys. Rev. B {\bf 39}, 11663 (1989).

\bibitem{scal}
N. Bulut and D. J. Scalapino, Phys. Rev. B {\bf 53}, 5149 (1996).

\bibitem{vilk}
Y. M. Vilk and A. M. S. Tremblay, J. Phys. I France {\bf 7}, 1309 (1997).

\bibitem{BLee}
J. Brinckmann and P. A. Lee, Phys. Rev. Lett. {\bf 82}, 2915 (1999)
and Phys. Rev. B {\bf 65}, 014502 (2001).

\bibitem{AMT}
A.-M. S. Tremblay, B. Kyung and D. Senechal, Low Temp. Phys. {\bf 32}, 424 (2006).

\bibitem{schrieff}
J. R. Schrieffer, {\it Theory of Superconductivity} (Benjamin/Cummings,
Reading, 1964).

\bibitem{pwa}
H. F. Fong, B. Keimer, P. W. Anderson, D. Reznik, F. Dogan, and I. A. Aksay,
Phys. Rev. Lett. {\bf 75}, 316 (1995).

\bibitem{millis}
L. B. Ioffe and A. J. Millis, Science {\bf 285}, 1241 (1999).

\bibitem{mike95}
M. R. Norman, M. Randeria, H. Ding, and J. C. Campuzano, Phys. Rev. B
{\bf 52}, 615 (1995).

\bibitem{wang}
C. Li, S. Zhou and Z. Wang, Phys. Rev. B {\bf 73}, 060501 (2006).

\bibitem{mike1}
M.R. Norman, Phys. Rev. B {\bf 61}, 14751 (2000).

\bibitem{mike2}
M.R. Norman, Phys. Rev. B {\bf 63}, 092509 (2001).

\bibitem{adam05}
A. Kaminski, H. M. Fretwell, M. R. Norman, M. Randeria, S. Rosenkranz, U. Chatterjee, 
J. C. Campuzano, J. Mesot, T. Sato, T. Takahashi, T. Terashima, M. Takano, K. Kadowaki,
Z. Z. Li and H. Raffy, Phys. Rev. B {\bf 71}, 014517 (2005).

\bibitem{bifoot}
The dispersion being fitted corresponds to the bonding (bilayer) one.

\bibitem{gang5}
I. Eremin, D. K. Morr, A.V. Chubukov, K. H. Bennemann and M. R. Norman,
Phys. Rev. Lett. {\bf 94}, 147001 (2005).

\bibitem{fauque}
B. Fauque, Y. Sidis, L. Capogna, A. Ivanov, K. Hradil, C. Ulrich, A. I. Rykov, B. Keimer and
P. Bourges, cond-mat/0701052.

\bibitem{tranq04} J.M. Tranquada, H. Woo, T. G. Perring, H. Goka, G. D. Gu, G. Xu,
M. Fujita and K. Yamada, Nature (London) {\bf 429}, 534 (2004).

\bibitem{hayd04} S.M. Hayden, H. A. Mook, P. Dai, T. G. Perring and F. Dogan,
Nature (London) {\bf 429}, 531 (2004).

\bibitem{zhou}
X. J. Zhou, T. Yoshida, D.-H. Lee, W. L. Yang, V. Brouet, F. Zhou, W. X. Ti, J. W. Xiong, Z. X. Zhao,
T. Sasagawa, T. Kakeshita, H. Eisaki, S. Uchida, A. Fujimori, Z. Hussain and Z.-X. Shen,
Phys. Rev. Lett. {\bf 92}, 187001 (2004).

\bibitem{bilayer}
A. Kaminski, S. Rosenkranz, H. M. Fretwell, M. R. Norman, M. Randeria, J. C. Campuzano,
J.-M. Park, Z. Z. Li, H. Raffy, Phys. Rev. B {\bf 73}, 174511 (2006).

\bibitem{eremin2}
I. Eremin, D. K. Morr, A.V. Chubukov, K. H. Bennemann, cond-mat/0611267.

\bibitem{schnyder}
A. P. Schnyder, A. Bill,  C.  Mudry, R. Gilardi, H. M. Ronnow, J. Mesot, Phys. Rev. B {\bf 70},
214511 (2004).

\bibitem{shulz}
H. J. Schulz, Phys. Rev. Lett. {\bf 64}, 1445 (1990).

\bibitem{pailhes}  S. Pailhes, Y. Sidis, P. Bourges, V. Hinkov, A. Ivanov, C. Ulrich, L. P. Regnault,
and B. Keimer, Phys. Rev. Lett.
{\bf 93}, 167001 (2004).

\bibitem{zhounat} X. J. Zhou, T. Yoshida, A. Lanzara, P. V. Bogdanov, S. A. Kellar, K. M. Shen,
W. L. Yang, F. Ronning, T. Sasagawa, T. Kakeshita, T. Noda,  H. Eisaki, S. Uchida, C. T. Lin,
F. Zhou, J. W. Xiong, W. X. Ti,  Z. X. Zhao, A. Fujimori, Z. Hussain and Z.-X. Shen,
Nature (London) {\bf 423}, 398 (2003).

\bibitem{utpal} U. Chatterjee, D. K. Morr, M. R. Norman, M. Randeria, A. Kanigel, M. Shi,
E. Rossi, A. Kaminski, H. M. Fretwell, S. Rosenkranz, K. Kadowaki, and J. C. Campuzano,
cond-mat/0606346 (2006).

\end{thebibliography}
\end{document}